\documentclass{article}
\usepackage{icrctc07}

\title{{Studies of clustering in the arrival directions of cosmic rays detected
at the Pierre Auger Observatory above 10 EeV
}}
\shorttitle{Studies of clustering}
\authors{Silvia Mollerach$^1$, for the Pierre Auger Collaboration}
\shortauthors{Pierre Auger Collaboration}
\afiliations{\it Observatorio Pierre Auger, av. San Mart\'{\i}n Norte 304, 
(5613)
  Malarg\"ue, Argentina\\
$^1$ Depto de F\'\i sica, Centro At\'omico Bariloche, CNEA and CONICET, 
Argentina}
\email{mollerach@cab.cnea.gov.ar}

\abstract{If clustering of the arrival directions of ultra high energy 
cosmic rays is discovered, this would provide important information 
about their origin, 
composition, and the galactic and extragalactic magnetic
fields. We present here the analysis of the
autocorrelation function of the data from the Pierre Auger Observatory
as a function of the angular scale and the energy threshold.
We compare our results with the signals found by
previous experiments.}

\begin{document}
\maketitle

\section{Introduction}

The identification of the sources of the ultra high energy cosmic rays 
is one of the main open problems in astrophysics. The study of their
arrival directions is likely to provide significant insight into this 
question. If cosmic rays are charged particles, their trajectories
will be bent by the intervening galactic and extragalactic magnetic 
fields and the arrival directions will not point back to their sources. 
The intensity and orientation of these fields are not well known, but as
the deflections decrease with the inverse of the energy, the effect is 
smaller at the largest energies. Thus, it is at the highest energies 
that cosmic rays are most likely to point towards their sources.

On the other hand, the distance from which ultra high energy protons 
can arrive to the Earth is expected to be limited by the energy loses 
caused by the photo-pion production processes
in the interaction with the cosmic microwave background (GZK effect 
\cite{Greisen:1966jv,Zatsepin:1966jv}), and similarly nuclei can 
undergo photo-disintegration processes. 
This  would strongly attenuate the flux coming from distant sources. 
Hence, at energies above $\sim 60$ EeV, 
cosmic rays are expected to come  mostly from nearby sources.

These ideas have motivated an extensive search of clustering signals both
at small angular scales, looking for point-like sources, and at 
intermediate angular scales, looking for the pattern characterizing the 
distribution of nearby sources. Although the data from a number of experiments
have shown a remarkably isotropic distribution of arrival directions, there
has been a claim of small scale clustering at energies larger than 
$40$ EeV by the AGASA experiment 
\cite{Hayashida:1996bc,Takeda:1999sg,Teshima:2003wd}. The HiRes 
experiment has found no significant clustering signal at any angular scale
up to $5^\circ$ for any energy above $10$ EeV \cite{Abbasi:2004ib}.
A hint of correlation at scales around $25^\circ$ and energies above 
$40$ EeV, combining data from HiRes  stereo, AGASA, Yakutsk and
SUGAR experiments has been pointed out in ref. \cite{Kachelriess:2005uf}.

\section{The data set}\label{dataset}

We use data recorded by the Surface Detector of the Pierre Auger 
Observatory between 1 January 2004 and 15 March 
2007 with energies above $10$ EeV and zenith angle smaller than $60^\circ$.
This represents a set of 1672 events, with 62 of them having energies larger 
than $40$ EeV, that pass our reconstruction quality cuts, 
with 5 active stations surrounding the station with 
the highest signal and with the reconstructed core position lying inside 
an active triangle of stations. 
We only considered events that triggered 6 or more stations, which
angular resolution, defined as the angular radius around the true cosmic ray
direction that would contain 
$68\%$ of the reconstructed shower directions, is at 
these energies  $0.9^\circ$ \cite{ave}. 
The energy is calibrated using the hybrid 
events simultaneously detected by the fluorescence and the surface detectors.

The fact that the surface detectors array is fully efficient for 
events with energy larger than 3 EeV implies that the exposure area
is determined only by geometric factors leading to a simple analytic 
dependence on declination. 
A small modulation in right ascension is also present due
to the growth of the array during the data taking period, but this effect 
is small and can be ignored in this analysis.

\section{Methods}\label{methods}

\begin{center}
  \begin{figure}[!ht]
      \begin{center}        
      \includegraphics[draft=false,scale=0.44,angle=-90]{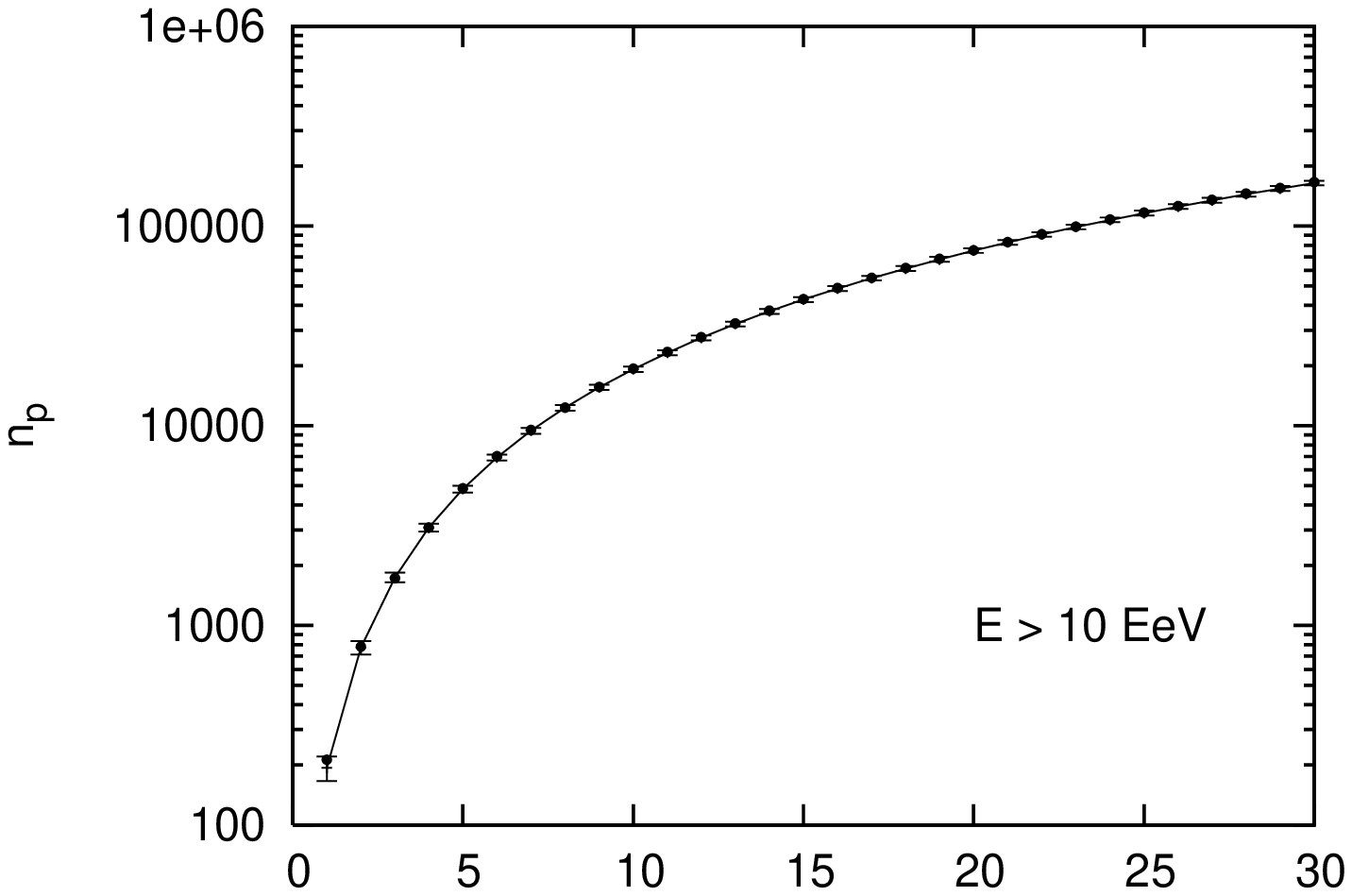}
      \includegraphics[draft=false,scale=0.43,angle=-90]{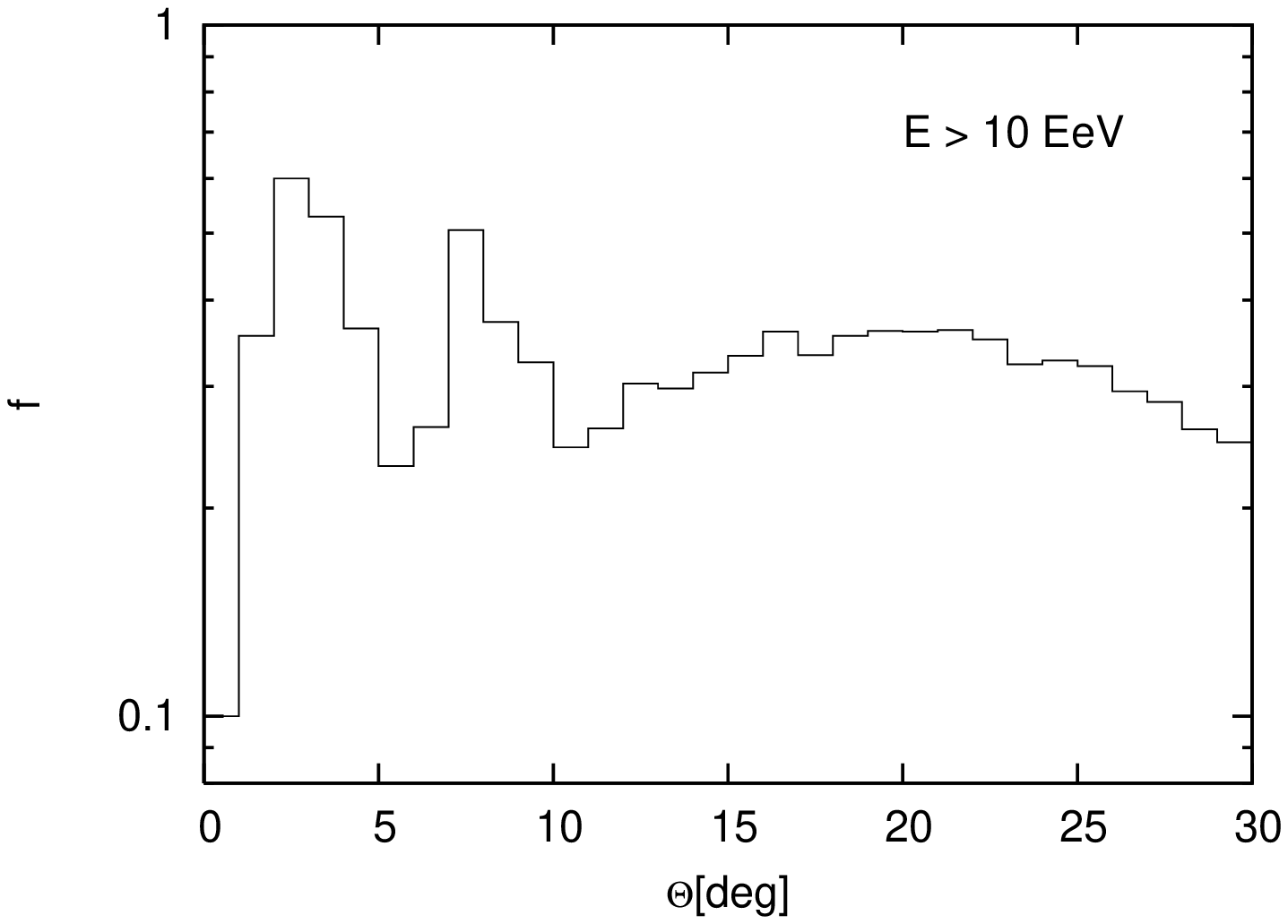}
\caption{Upper panel: Autocorrelation function above 10 EeV as a function 
of the angle (dots) and autocorrelation function for an isotropic 
distribution with $95\%$ confidence level band.
Lower panel: Fraction of isotropic simulations with larger number of 
pairs than the data.
}\label{corre.a10}
      \end{center}
  \end{figure}
\end{center}

A standard tool for studying anisotropies is the two-point angular 
correlation function. This counts the number of pairs separated by less 
than an angle $\theta$ among the events with energy larger than an energy 
threshold $E$. We show in Figure \ref{corre.a10} the result for all the events 
above 10 EeV. The expected number of pairs is obtained by generating a 
large number of Monte Carlo simulations with the same number of events 
with an isotropic distribution modulated by the exposure of the detector,
from which the mean number of pairs and the $95\%$ CL band is extracted for
each angular scale. The fraction of simulations with a larger number of 
pairs than the data gives a measure of the probability that an observed 
excess of pairs arise by chance from an isotropic distribution of events.

The result of the autocorrelation function analysis 
depends of course on the values of $\theta$ and $E$ considered.
However, the fact that the deflections expected from galactic and 
extragalactic magnetic fields and the distribution of the sources are largely
unknown makes it difficult to fix these values a priori. The significance 
of an autocorrelation signal at a given angle and energy when 
these values have not been fixed a priori is a delicate issue, that
has made the AGASA small scale clustering claim very controversial
\cite{Finley:2003ur}.

\begin{figure*}[th]
\begin{center}
\includegraphics [width=0.70\textwidth]{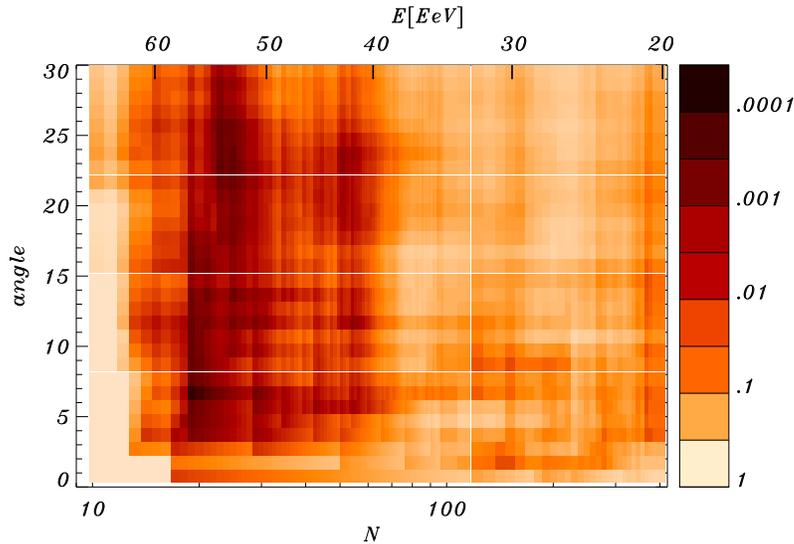}
\end{center}
\caption{Autocorrelation scan for events with energy above 20 EeV}\label{scan}
\end{figure*}

We adopt here the method proposed by Finley and Westerhoff 
\cite{Finley:2003ur}, 
in which a scan over the minimum energy and the angle is performed.
For each value of $E$ and $\theta$ a chance probability is estimated 
by generating a large number of isotropic Monte Carlo simulations of the same 
number of events, and computing the fraction of simulations 
having an equal or larger number of pairs than the data for those values 
of $E$ and $\theta$. The most relevant clustering signal corresponds 
to the values of $\theta$ and $E$ that have the smaller value of the 
chance probability, $P_{min}$. Finally, the probability that such 
clustering arises by chance from an isotropic distribution is estimated 
by performing a similar scan on a large number of
isotropic data sets simulated by the Monte Carlo technique
and finding the fraction of the simulations having a
smaller $P_{min}$ than the data.

We show in Figure \ref{scan} the result of the scan above a minimum
energy of $20$ EeV and up to a maximum angle of $30^\circ$.
A broad region with an excess of correlation appears at intermediate angular 
scales and large energies. 
The minimum appears at $7^\circ$ for the
19 highest energy events ($E > 57.5$ EeV), where 8 pairs are observed, while
1 was expected. 
The fraction of isotropic simulations with larger number of pairs at that 
angular scale and for that number of events is  $P_{min}=10^{-4}$, obtained by 
comparing the observed number of pairs with that arising in $10^6$ 
isotropic simulations.
An extended scan for the 1672 events with $E > 10$ EeV shows no new minimum.

The chance probability of a $P_{min} < 10^{-4}$ to
arise from an isotropic distribution, obtained by performing the
same scan to $10^5$ simulations,  is $P \simeq 2\times 10^{-2}$.

Previous analyses of other experiments have reported small scale
clustering signals at $2.5^\circ$ in AGASA data 
\cite{Hayashida:1996bc,Takeda:1999sg,Teshima:2003wd}
and at intermediate scales (around $25^\circ$) in a combination 
of data from different experiments \cite{Kachelriess:2005uf}, 
both for energies above $40$ EeV. 
\begin{center}
  \begin{figure}[!ht]
      \begin{center}        
      \includegraphics[draft=false,scale=0.44,angle=-90]{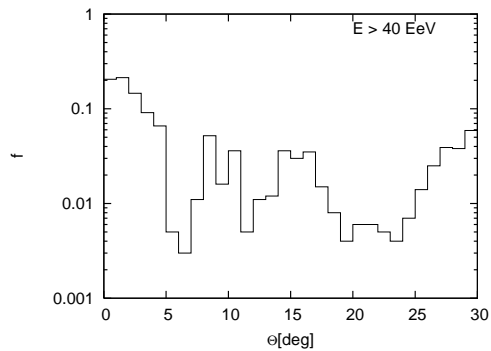}
\caption{Fraction of simulations with larger number of pairs than the
  data for the 64 events with $E > 40$ EeV. 
}\label{pr40}
      \end{center}
  \end{figure}
\end{center}

\begin{center}
  \begin{figure}[!ht]
      \begin{center}        
      \includegraphics[draft=false,scale=0.44,angle=-90]{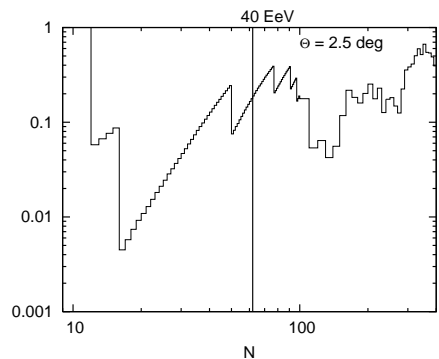}
\caption{Fraction of simulations with more pairs separated by  
$\theta < 2.5^\circ$, as a function of the number of events.
}\label{pr2.5}
      \end{center}
  \end{figure}
\end{center}

We show in Figure \ref{pr40} the fraction of simulations with more pairs 
than the data for the events
with $E > 40$ EeV (62 events). The small scale clustering in Auger data
is compatible with that expected from an isotropic flux with our present
statistics. We observe 2 pairs within $2.5^\circ$, 
while 1.5 were expected from an isotropic flux.
Due to a possible difference in the energy calibration between
Auger and AGASA, the clustering signal reported by AGASA could appear 
in Auger data at a different (lower) energy scale.  
We show in Figure \ref{pr2.5} the probability
for a larger or equal number of pairs within  $\theta = 2.5^\circ$ as a 
function of the number of events (or threshold energy). 
At this angular scale, no strong excess of clustering appears in the present 
data set. In the relevant energy range there is a slight excess of pairs, 
for example for $N=150$ ($E > 30$ EeV) 14 pairs are observed 
while 8.5 are expected, with a 
probability for this to happen by chance in an isotropic distribution of about
5\%. This excess is anyhow much smaller than the one reported by AGASA, where 
7 pairs were observed while 1.45 were expected (out of 57 events).

Regarding the intermediate angular scales signal, some hint of clustering,
though not very significant with the present statistics, is apparent.
It is weaker at this energy than at higher energies (above 50 EeV) as 
discussed above (Figure \ref{scan}).

\section{Conclusion}\label{conclusion}

We have searched for clustering signals in the Auger data with energies 
above 10 EeV. In particular we have checked the clustering signal at
$2.5^\circ$ for $E > 40$ EeV reported by AGASA. No strong excess
of clustering is present at this angular scale in our data set.

An extensive scan in angle and energy threshold shows some hints of 
clustering at larger energies ($E > 50$ EeV) and intermediate angular scales, 
that could be a signal of the large scale distribution of nearby sources. 
However, taking 
into account the scan performed, the probability that this kind of signals 
appear by chance from an isotropic flux is $P = 2\%$. Thus it is only 
marginally significant with our present statistics. 
Auger future data will be used to check if this correlation is
real.

\bibliography{icrc0074}
\bibliographystyle{unsrt}

\end{document}